\documentclass[%
 reprint,
superscriptaddress,
 amsmath,amssymb,
 aps,
pra,
]{revtex4-1}

\usepackage{amsthm,amsmath,amssymb}
\usepackage{mathrsfs}
\usepackage{amsthm,amssymb}
\usepackage{amsmath}
\usepackage{graphicx}
\usepackage{dcolumn}
\usepackage{bm}
\usepackage{multirow} 
\usepackage{mathdots}
\usepackage{enumerate}
\usepackage{algorithm}
\usepackage{algorithmic}
\usepackage{color}
\usepackage{epsfig}
\usepackage{subfigure}
\usepackage{caption}
\usepackage[greek,english]{babel}

\theoremstyle{remark}

\UseRawInputEncoding

\begin{document}

\makeatletter
\newcommand{\ud}{\mathrm{d}}
\newcommand{\rmnum}[1]{\romannumeral #1}
\newcommand{\Rmnum}[1]{\expandafter\@slowromancap\romannumeral #1@}
\newcommand{\udots}{\mathinner{\mskip1mu\raise1pt\vbox{\kern7pt\hbox{.}}
        \mskip2mu\raise4pt\hbox{.}\mskip2mu\raise7pt\hbox{.}\mskip1mu}}
\makeatother

\preprint{APS/123-QED}

\title{Quantum Algorithm for Anomaly Detection of Sequences}
\author{Ming-Chao Guo}
\affiliation{State Key Laboratory of Networking and Switching Technology, Beijing University of Posts and Telecommunications, Beijing, 100876, China}
\affiliation{State Key Laboratory of Cryptology, P.O. Box 5159, Beijing, 100878, China}
\author{Hai-Ling Liu}
\affiliation{State Key Laboratory of Networking and Switching Technology, Beijing University of Posts and Telecommunications, Beijing, 100876, China}
\author{Shi-Jie Pan}
\affiliation{State Key Laboratory of Networking and Switching Technology, Beijing University of Posts and Telecommunications, Beijing, 100876, China}
\author{Wen-Min Li}
\email{liwenmin@bupt.edu.cn}
\author{Su-Juan Qin}
\affiliation{State Key Laboratory of Networking and Switching Technology, Beijing University of Posts and Telecommunications, Beijing, 100876, China}
\author{Xin-Yi Huang}
\affiliation{Artificial Intelligence Thrust, Information Hub, Hong Kong University of Science and Technology, Guangzhou, 511455, China}
\author{Fei Gao}
\email{gaof@bupt.edu.cn}
\affiliation{State Key Laboratory of Networking and Switching Technology, Beijing University of Posts and Telecommunications, Beijing, 100876, China}
\author{Qiao-Yan Wen}
\affiliation{State Key Laboratory of Networking and Switching Technology, Beijing University of Posts and Telecommunications, Beijing, 100876, China}
\date{\today}

\begin{abstract}
Anomaly detection of sequences is a hot topic in data mining. Anomaly Detection using Piecewise Aggregate approximation in the Amplitude Domain (called ADPAAD) is one of the widely used methods in anomaly detection of sequences. The core step in the classical algorithm for performing ADPAAD is to construct an approximate representation of the subsequence, where the elements of each subsequence are divided into several subsections according to the amplitude domain and then the average of the subsections is computed. It is computationally expensive when processing large-scale sequences. In this paper, we propose a quantum algorithm for ADPAAD, which can divide the subsequence elements and compute the average in parallel. Our quantum algorithm can achieve polynomial speedups on the number of subsequences and the length of subsequences over its classical counterpart.
\end{abstract}

\pacs{Valid PACS appear here}
\maketitle

\section{Introduction}
Anomaly detection refers to the problem of finding patterns in data that do not conform to expected behavior \cite{VAV2009}. It can be divided into three modes according to the availability of data labels: supervised anomaly detection, semi-supervised anomaly detection, and unsupervised anomaly detection \cite{VAV2009}. Anomaly Detection using Piecewise Aggregate approximation in the Amplitude Domain (called ADPAAD) \cite{RLL2017} is important unsupervised anomaly detection. It has been used in various fields, as a widely used method in anomaly detection of sequences, such as medical science \cite{TTP2011}, network security \cite{VDM2009}, finance \cite{CFL2011}, industrial engineering \cite{ADK2010}, and transportation \cite{LBF2013}.  

The classical algorithm of ADPAAD can be divided into the following parts. (1) Division: the sequence is divided into subsequences through a sliding window. (2) Approximate representation: the elements of each subsequence are divided into several subsections according to the amplitude domain, and then the average of the subsections is computed to construct an approximate representation of the subsequence. (3) Similarity: the similarity between subsequences is calculated according to their approximate representation. (4) Anomaly score: anomaly score for each subsequence is calculated. (5) Determination: anomalous subsequences are determined based on their anomaly score.
Among them, obtaining the approximate representation of subsequences typically carries a linear time overhead with respect to the input size, and the running time of computing similarity depends quadratically on the number of subsequences. These are computationally expensive when processing large-scale sequences.

Quantum computing has been shown to be more computationally powerful over classical computing in solving certain problems, such as factoring integers \cite{PW1994}, searching in unstructured databases \cite{LK1996}, solving systems of linear \cite{AAS2009,LCS2018} and differential \cite{HYW2021} equations, cryptanalysis \cite{ZBH2022,XZX2019}, and private queries \cite{CXT2020,FSW2019,GLM2008}. The combination of quantum computing and machine learning has made great progress in classification \cite{SMP2013,NDS2012}, clustering \cite{SBS2017}, neural networks \cite{PTC2018}, linear regression \cite{GM2017,CFQ2019,CFC2019}, association rule mining \cite{CFQ2016}, dimensionality reduction \cite{IL2016,SMP2014,SLH2020,CFS2019}, and quantum support vector machine \cite{PMS2014,ZLH2020}, etc. Therefore, it is worthwhile to explore quantum algorithms for anomaly detection to reduce its computational complexity.

Several works have been developed in the context of quantum computing to solve anomaly detection problems. In 2018, Liu et al. proposed a quantum kernel principal component analysis algorithm for anomaly detection \cite{NP2018}. 
It achieves exponential speedup on the dimension of the training data set. Subsequently, Liang et al. presented a quantum anomaly detection algorithm based on density estimation \cite{JSM2019}. Its complexity is logarithmic in the dimension and the number of training data compared to the corresponding classical algorithm. In 2022, Guo et al. proposed a quantum algorithm for anomaly detection \cite{MHY2021}, which achieves exponential speedup on the number of training data points over its classical counterpart. These quantum algorithms are aimed at semi-supervised anomaly detection and cannot be directly applied to ADPAAD.


In this paper, we focus on studying quantum algorithms for unsupervised anomaly detection. Specifically, we propose a quantum algorithm for ADPAAD. As show above, the core step in the classical algorithm for performing ADPAAD is the approximate representation. To reduce computational complexity, quantum multiply-adder \cite{LJ2017,STJ2017} is used to divide the subsequence elements into several subsections, and amplitude amplification and estimation \cite{GPM2002} are used to calculate the average of each subsection without counting the number of elements belonging to the same subsection. This enables both steps to be implemented in parallel. In practical application scenarios, due to the huge amount of data and the difficulty of collecting abnormal labels or normal label samples, the data is often unlabeled. However, there is currently no quantum algorithm specifically proposed for unsupervised anomaly detection. Our quantum algorithm achieves polynomial speedups compared to its classical counterpart.

An outline of the paper follows. In Sec.~\ref{sec2}, we review the related definitions of anomaly detection and briefly introduce the classical algorithm of ADPAAD. In Sec.~\ref{sec3}, we propose a quantum algorithm for ADPAAD and analyze its complexity in detail. The conclusion is given in Sec.~\ref{sec4}. Finally, we provide a detailed analysis of the general case of step 1 in Appendix~\ref{sec6}.

\section{Review of ADPAAD}
\label{sec2}

In this section, we introduce the relevant definitions for anomaly detection of sequences used in this paper and briefly review the classical algorithm of ADPAAD \cite{RLL2017}.
\subsection{Definitions}

For convenience, we begin with the following related definitions \cite{RLL2017}:

\textbf{Definition 1}. Sequence: A sequence $X(m)=\{x(1),x(2),\cdots,x(m)\}$ is a time series, where data elements are sorted by time, and $m$ represents the length of $X(m)$.

\textbf{Definition 2}. Sliding window: A user-defined window of length $n\le m$, all possible subsequences can be extracted by sliding a window of size $n$ across the sequence $X(m)$.

\textbf{Definition 3}. Subsequence: Given a sequence $X(m)$, the subsequence of length $n$ is extracted through a sliding window. The $k$-th subsequence can be expressed as
\begin{equation}
X_{k}=\{x_{k}(1),x_{k}(2),\cdots,x_{k}(n)\}.
\end{equation}

\textbf{Definition 4}. Amplitude domain: Given a subsequence $X_{k}$, its amplitude domain is defined as $I_{k}=[L_{k},H_{k}]$, where $L_{k}$ and $H_{k}$ represent the minimum and maximum values of the $X_{k}$, respectively.

\textbf{Definition 5}. Subsection: Given a subsequence $X_{k}$, its subsections are generated by dividing the amplitude domain $I_{k}$. The $t$-th subsection of the subsequence $X_{k}$ can be shown as follows:
\begin{equation}
I_{k}^{t}=[a_{k}^{t-1},  a_{k}^{t}),~~ 1\le t\le q,
\end{equation}
where $a_{k}^{t-1}$ and $a_{k}^{t}$ denote the lower and upper bounds of the $t$-th subsection, respectively. The $q$ $(q\ll n)$ represents the number of subsections.
\subsection{Classical Algorithm of ADPAAD }

An important step of implementing the ADPAAD is a piecewise aggregated approximate representation in the amplitude domain (called PAAD representation) for each subsequence, which reduces dimensionality of the subsequence and preserves its key information. In this step, the amplitude domain of each subsequence is determined according to Definition 4, and divided into $q$ subsections according to Definition 5. 
If the element value of the subsequence $X_{k}$ is within $[a_{k}^{t-1},  a_{k}^{t})$, then it is assigned to the $t$-th subsection. The representation of subsequence $X_{k}$ is written as a vector of the mean elements in subsections. The whole procedure is depicted as follows.

(a) Divide the sequence $X(m)$ into subsequences $X_{1},X_2,\cdots,X_{K}$ as shown in Definition 3 by sliding windows.

(b) Construct a PAAD representation for each subsequence $X_{k}$ to get
\begin{equation}
\bar{X}_{k}=[\mu_{k}^1,\mu_{k}^2,\cdots,\mu_{k}^q]^{T}, ~~\mu_{k}^{t}=\frac{1}{n_{k}^t}\sum_{x_{k}\in[a_{k}^{t-1},  a_{k}^{t})}x_{k}^{t}.
\end{equation}
where $k=1,2,\cdots,K$ and $n_{k}^t$ denotes the number of data points belonging to the subsection $[a_{k}^{t-1},  a_{k}^{t})$ in subsequence $X_{k}$.

(c) The similarity between $X_{i}$ and $X_{j}$ is defined in terms of the Euclidean distance as follows:
\begin{equation}
S_{\mu}(X_i,X_j)=\sqrt{\sum_{t=1}^{q}(\mu_{i}^{t}-\mu_{j}^{t})^2}.
\end{equation}

(d) The anomaly score of the $i$-th subsequence $X_i$ can be calculated by
\begin{equation}
h_{i}=\frac{\sum_{j=1}^{K}S_{\mu}(X_i,X_j)/K}{\sum_{i=1}^{K}\sum_{j=1}^{K}S_{\mu}(X_i,X_j)/K^2}.
\end{equation}

(e) Set a threshold $\delta$ in advance, if $h_{i}\geq\delta$, we mark the subsequence as an anomaly; otherwise, it is judged as normal.

The total runtime of this algorithm is $O(Kqn+K^2q)$. Nevertheless, in the current era of big data, processing large-scale sequences results in huge time complexity. That is why a quantum algorithm for ADPAAD is needed.
\section{Quantum Algorithm}
\label{sec3}
In this section, we present a quantum algorithm for ADPAAD and analyze its complexity in detail.

Our quantum algorithm consists of four steps, corresponding to the four parts (b)-(e) of the classical ADPAAD algorithm. In step 1, we prepare the quantum state $\frac{1}{\sqrt{K}}\sum_{i=1}^{K}|i\rangle\frac{1}{\sqrt{q}}\sum_{t=1}^{q}|t\rangle|\mu_{i}^{t}\rangle$ for the PAAD representation of each subsequence; the quantum state $\frac{1}{K}\sum_{i,k=1}^{K}|i\rangle|k\rangle|S_{\mu}(X_{i},X_{k})\rangle$ is generated in step 2. In step 3, we perform amplitude estimation \cite{GPM2002} to get the state $\frac{1}{\sqrt{K}}\sum_{i=1}^{K}|i\rangle|h_{i}\rangle$. Step 4 performs Grover's algorithm to search for anomalous subsequences satisfying $h_{i}\geq\delta$. The entire algorithm process is shown in Fig. 1.
\begin{figure*}[htb]
	\centering
		\includegraphics[height=10cm,width=11cm]{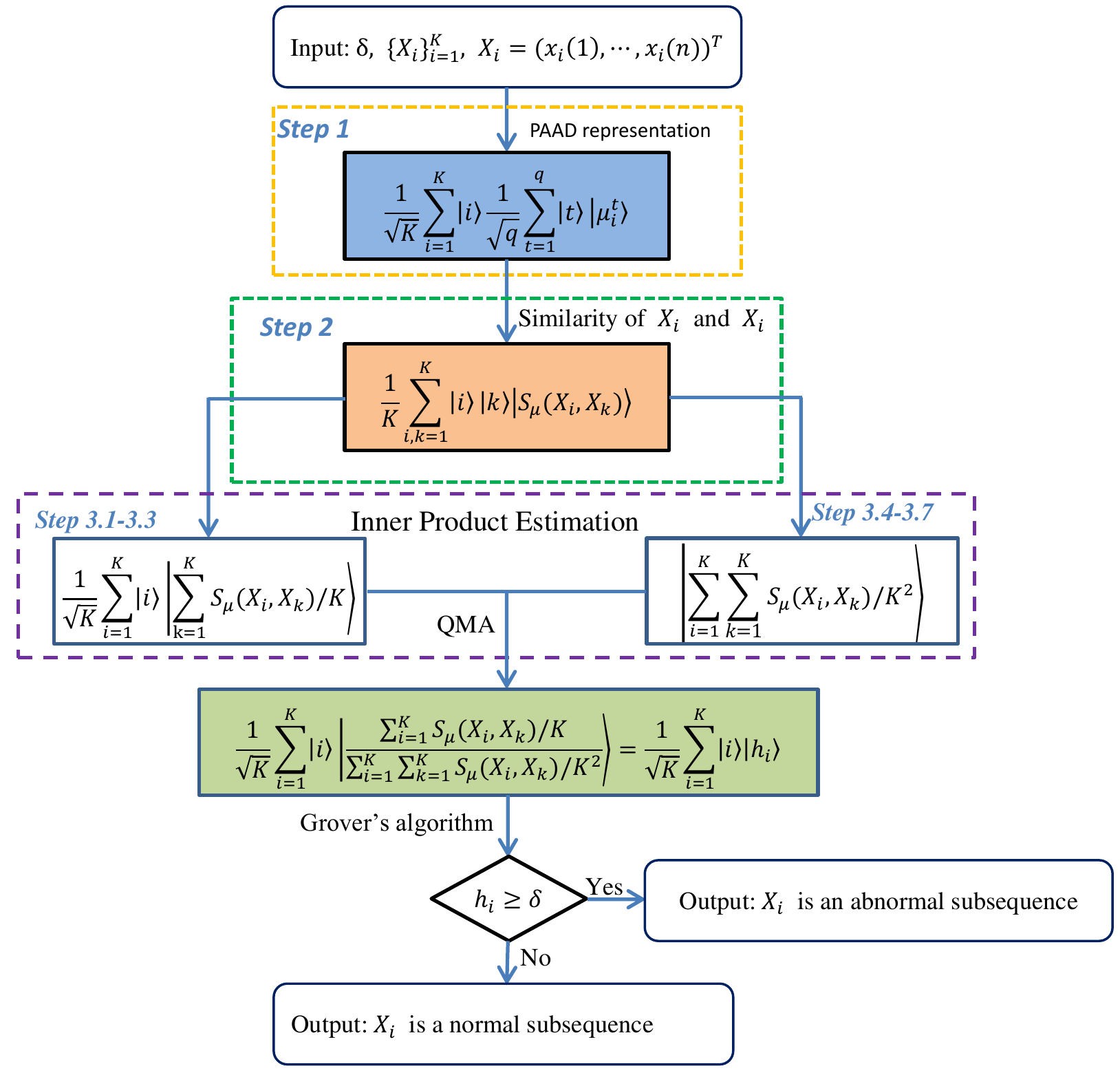}
\caption{The process diagram of quantum ADPAAD algorithm, where $\delta$ is a predetermined threshold and QMA denotes quantum multiply-adder.}
	\label{fig:kappa}
\end{figure*}

\subsection{Preliminaries}

The sequence $X(m)$ is divided into subsequences $X_{1},X_{2},\cdots,X_{K}$ as shown in Definition 3 by sliding windows.

\textbf{Lemma 3.1} (QRAM \cite{VSL2008}). Assume that the subsequences $\{X_{i}\}_{i=1}^{K}$, upper and lower bounds of subsections (that is, $a_i^t$ and $a_i^{t-1}$, respectively) are stored in a Quantum Random Access Memory (QRAM) \cite{VSL2008}, which allows us to efficiently perform the following two unitary operations in $O(log Kn)$ and $O(log Kq)$ time as given below:
\begin{align}
&O_{X}:|i\rangle|j\rangle|0\rangle\rightarrow|i\rangle|j\rangle|x_i(j)\rangle,\nonumber \\
&O_{s}:|i\rangle|t\rangle|0\rangle|0\rangle\rightarrow|i\rangle|t\rangle|a_{i}^t\rangle|a_{i}^{t-1}\rangle,
\end{align}
where $i=1,2,\cdots,K,~j=1,2,\cdots,n$ and $t=1,2,\cdots q$.

The QRAM has been used to process data classification \cite{SMP2013,NDS2012}, clustering \cite{SBS2017}, neural networks \cite{PTC2018}, linear regression \cite{GM2017,CFQ2019,CFC2019}, association rule mining \cite{CFQ2016}, dimensionality reduction \cite{IL2016,SMP2014}, and other state preparation tasks in quantum algorithms. However, how to physically build QRAM is still an open question, and it may be a goal worthy of consideration in the future.


\textbf{Lemma 3.2} (Inner Products Estimation \cite{KJA2019}). Assume that there are unitaries $|i\rangle|0\rangle\rightarrow|i\rangle|\bm{v}_i\rangle$, and $|j\rangle|0\rangle\rightarrow|j\rangle|\bm{w}_j\rangle$ can
be performed in time $T$ and the norms of the vectors are known. For any $\epsilon>0$, there exists a quantum algorithm that can estimate $\langle \bm{v}_i|\bm{w}_j\rangle$ in time $O(\frac{||\bm{v}_i||||\bm{w}_j||T}{\epsilon})$ to obtain the state $|i\rangle|j\rangle|0\rangle\rightarrow|i\rangle|j\rangle|\overline{\langle \bm{v}_i|\bm{w}_j\rangle}\rangle$, where $|\overline{\langle \bm{v}_i|\bm{w}_j\rangle}-\langle \bm{v}_i|\bm{w}_j\rangle|<\epsilon$.

\subsection{Algorithm}
The specific process of our quantum algorithm is as follows:
\subsubsection{\textbf{Prepare the state $\frac{1}{\sqrt{K}}\sum_{i=1}^K|i\rangle\frac{1}{\sqrt{q}}\sum_{t=1}^{q}|t\rangle|\mu_{i}^{t}\rangle$.}}
We design a quantum algorithm to prepare the quantum state corresponding to the PAAD representation of each subsequence, and the specific process is shown as follows:

\textit{Step 1.1} Initializing the quantum state
\begin{equation}
|0^{\otimes\log K}\rangle_1|0^{\otimes\log q}\rangle_2|0\rangle_3|0\rangle_4|0^{\otimes\log n}\rangle_5|0\rangle_6|0\rangle_7,
\end{equation}
where the subscript numbers denote different registers.

\textit{Step 1.2} Perform the Hadamard gates $H^{\otimes\log K}$, $H^{\otimes\log q}$, and $H^{\otimes\log n}$ to obtain
\begin{equation}
\frac{1}{\sqrt{K}}\sum_{i=1}^K|i\rangle_1\frac{1}{\sqrt{q}}\sum_{t=1}^{q}|t\rangle_2|0\rangle_3|0\rangle_4\frac{1}{\sqrt{n}}\sum_{j=1}^{n}|j\rangle_5|0\rangle_6|0\rangle_7.
\end{equation}

\textit{Step 1.3} Apply the oracles $O_{X}$ and $O_{s}$, which can be seen in Eq. (6), to prepare
\begin{equation}
\frac{1}{\sqrt{Kqn}}\sum_{i=1}^K|i\rangle_1\sum_{t=1}^{q}|t\rangle_2|a_{i}^{t-1}\rangle_3|a_{i}^{t}\rangle_4\sum_{j=1}^{n}|j\rangle_5|x_i(j)\rangle_6|0\rangle_7.
\end{equation}

\textit{Step 1.4} Perform the quantum multiply-adder (QMA) gate \cite{LJ2017,STJ2017} on the third, fourth, and sixth registers, and undo the redundant registers to create
\begin{equation}
\frac{1}{\sqrt{K}}\sum_{i=1}^K|i\rangle_1\frac{1}{\sqrt{q}}\sum_{t=1}^{q}|t\rangle_2\frac{1}{\sqrt{n}}\sum_{j=1}^{n}|j\rangle_5|x_i(j)\rangle_6|\rho_{i}^{t}(j)\rangle_7,
\end{equation}
where $\rho_{i}^{t}(j)=(x_{i}(j)-a_i^{t-1})(x_{i}(j)-a_i^t)$. If $\rho_{i}^{t}(j)\leq0$, then we consider $x_i(j)\in I_{i}^t=[a_i^{t-1},a_i^t)$, otherwise, $x_i(j)\notin [a_i^{t-1},a_i^t)$. That is, Eq. (11) can be rewritten as:
\begin{align}
&\frac{1}{\sqrt{Kq}}\sum_{i=1}^K|i\rangle\sum_{t=1}^{q}|t\rangle\frac{1}{\sqrt{n}}[\sum_{x_{i}(j)\in I_{i}^{t}}|j\rangle|x_i(j)\rangle|\rho_{i}^{t}(j)\leq0\rangle\nonumber\\
&+\sum_{x_{i}(j)\notin I_{i}^{t}}|j\rangle|x_i(j)\rangle|\rho_{i}^{t}(j)>0\rangle].
\end{align}

\textit{Step 1.5} Execute the amplitude amplification \cite{GPM2002,SLH2022} to obtain
\begin{equation}
\frac{1}{\sqrt{K}}\sum_{i=1}^K|i\rangle_1\frac{1}{\sqrt{q}}\sum_{t=1}^{q}|t\rangle_2(\sqrt{p}|\Phi_i^t\rangle+\sqrt{1-p}|(\Phi_i^t)^{\perp}\rangle)_{567},
\end{equation}
where $|\Phi_i^t\rangle=\frac{1}{\sqrt{n_{i}^{t}}}\sum_{x_{i}(j)\in I_{i}^{t}}|j\rangle|x_i(j)\rangle|\rho_{i}^{t}(j)\rangle$, $n_{i}^t$ denotes the number of data points belonging to the subsection $I_{i}^t$ in subsequence $X_{i}$. $p$ represents the probability of successfully measuring $|\Phi_i^t\rangle$ and $|(\Phi_i^t)^{\perp}\rangle$ is the quantum state that is orthogonal to $|\Phi_i^t\rangle$. For simplicity, we assume $p=1$, then Eq. (12) can be rewritten as
\begin{equation}
\frac{1}{\sqrt{K}}\sum_{i=1}^K|i\rangle_1\frac{1}{\sqrt{q}}\sum_{t=1}^{q}|t\rangle_2\frac{1}{\sqrt{n_i^t}}\sum_{x_{i}(j)\in I_{i}^{t}}|j\rangle_{5}|x_i(j)\rangle_{6}|\rho_{i}^{t}(j)\rangle_{7}.
\end{equation}
The general case is analyzed in detail in the Appendix.

\textit{Step 1.6} Appending one qubit and rotating it from $|0\rangle$ to $(\sqrt{\frac{x_{i}(j)}{C}}|0\rangle+\sqrt{1-\frac{x_{i}(j)}{C}}|1\rangle)$ controlled on $|x_i(j)\rangle$ \cite{AAS2009,BJY2018,KMK2019}, discard the sixth and seventh registers, we obtain the state
\begin{align}
&\frac{1}{\sqrt{Kq}}\sum_{i=1}^K|i\rangle_1\sum_{t=1}^{q}|t\rangle_2\frac{1}{\sqrt{n_{i}^{t}}}\sum_{x_{i}(j)\in I_{i}^{t}}|j\rangle_5\big(\sqrt{\frac{x_{i}(j)}{C}}|0\rangle+\nonumber\\
&\sqrt{1-\frac{x_{i}(j)}{C}}|1\rangle\big)_{8}=\frac{1}{\sqrt{K}}\sum_{i=1}^K|i\rangle_1\frac{1}{\sqrt{q}}\sum_{t=1}^{q}|t\rangle_2|\psi_{i}^{t}\rangle_{58},
\end{align}
where $C=\max|x_{i}(j)|$. The state $|\psi_{i}^{t}\rangle$ can be rewritten as $|\psi_{i}^{t}\rangle=\sin\theta_{i}^{t}|(\psi_{i}^{t})^0\rangle+\cos\theta_{i}^{t}|(\psi_{i}^t)^1\rangle$, where $|(\psi_{i}^{t})^0\rangle$ and $|(\psi_{i}^{t})^1\rangle$ represent the normalized quantum states of $\sqrt{\frac{x_{i}(j)}{C}}|j\rangle|0\rangle$ and $\sqrt{1-\frac{x_{i}(j)}{C}}|j\rangle|1\rangle$, respectively. It can be easily calculated: $\sin^2(\theta_{i}^{t})=\frac{1}{n_{i}^t}\sum_{x_{i}(j)\in I_{i}^t}\frac{x_{i}(j)}{C}$. We define $Q_{i}^t=-A_{i}^{t}S_0(A_{i}^t)^{\dagger}S_{\chi}$, where $A_{i}^t:|0\rangle_{58}\rightarrow|\psi_{i}^t\rangle,~S_0=I-2|0\rangle_{58}\langle0|_{58},~S_{\chi}=(I-2|0\rangle_{5}\langle0|_{5})\otimes I_8$. Perform $(Q_{i}^t)^l$ on the state $|\psi_{i}^t\rangle$ to get
$$(Q_{i}^t)^l|\psi_{i}^t\rangle=\sin[(2l+1)\theta_{i}^t]|(\psi_{i}^t)^0\rangle+\cos[2l+1)\theta_{i}^t](\psi_{i}^t)^1\rangle,$$
for any $l\in N$, $Q_{i}^t$ acts as a rotation in 2-dimensional space Span $\{|(\psi_{i}^t)^0\rangle,|(\psi_{i}^t)^1\rangle\}$, and it has two eigenvalues $e^{\pm\iota2\theta_{i}^t}$ with the eigenstates $|(\psi_{i}^t)^{\uparrow,\downarrow}\rangle$.

\textit{Step 1.7} Add an ancilla register and perform amplitude estimation on Eq. (14) to generate
\begin{equation}
\frac{1}{\sqrt{K}}\sum_{i=1}^K|i\rangle_1\frac{1}{\sqrt{q}}\sum_{t=1}^{q}|t\rangle_2(|(\psi_{i}^t)^{\uparrow}\rangle_{58}|\frac{\theta_{i}^t}{\pi}\rangle_9+|(\psi_{i}^t)^{\downarrow}\rangle_{58}|1-\frac{\theta_{i}^t}{\pi}\rangle_9).
\end{equation}

\textit{Step 1.8} Compute $|\mu_i^t\rangle=|C\cdot\sin^2(\theta_i^t)\rangle$ via the QMA and sine gate \cite{LJ2017,STJ2017}, undo redundant registers to get
\begin{equation}
\frac{1}{\sqrt{K}}\sum_{i=1}^K|i\rangle_1\frac{1}{\sqrt{q}}\sum_{t=1}^{q}|t\rangle_2|\mu_i^t\rangle_9.
\end{equation}
\subsubsection{\textbf{Prepare the state $\frac{1}{K}\sum_{i,k=1}^K|i\rangle|k\rangle|S_{\mu}(X_i,X_k)\rangle$.}}
We prepare quantum states corresponding to the similarity between subsequences, and the specific steps are as follows:

\textit{Step 2.1} Repeat the operations of step 1 to get
\begin{equation}
\frac{1}{K}\sum_{i,k=1}^K|i\rangle|k\rangle\frac{1}{\sqrt{q}}\sum_{t=1}^{q}|t\rangle|\mu_i^t\rangle|\mu_k^t\rangle.
\end{equation}

\textit{Step 2.2} Add an ancilla register and perform the QMA gate, uncompute the fourth and fifth registers to obtain
\begin{equation}
\frac{1}{K}\sum_{i,k=1}^K|i\rangle|k\rangle\frac{1}{\sqrt{q}}\sum_{t=1}^{q}|t\rangle|\mu_i^t-\mu_k^t\rangle.
\end{equation}

\textit{Step 2.3} Append an ancilla register and apply controlled rotation, we have the state
\begin{equation}
\frac{1}{K}\sum_{i,k=1}^K|i\rangle|k\rangle\frac{1}{\sqrt{q}}\sum_{t=1}^{q}|t\rangle[\frac{\mu_i^t-\mu_k^t}{2C}|0\rangle+\sqrt{1-(\frac{\mu_i^t-\mu_k^t}{2C})^2}|1\rangle].
\end{equation}

\textit{Step 2.4} Perform the amplitude estimation and compute $|\bar{S}_{\mu}(X_i,X_k)\rangle=|\sin(\alpha_{i,k})\rangle$ via the QMA and sine gate (similar to steps 1.7-1.8) to generate
\begin{equation}
\frac{1}{K}\sum_{i,k=1}^K|i\rangle|k\rangle|\bar{S}_{\mu}(X_i,X_k)\rangle,
\end{equation}
where $\bar{S}_{\mu}(X_i,X_k)=\frac{1}{\sqrt{q}}\sqrt{\sum_{t=1}^q(\frac{\mu_i^t-\mu_k^t}{2C})^2}=\frac{S_{\mu}(X_i,X_k)}{\sqrt{q}2C}$.\\
\subsubsection{\textbf{Prepare the quantum state $\frac{1}{\sqrt{K}}\sum_{i=1}^K|i\rangle|h_i\rangle$.}}
Demonstrating the process of obtaining anomaly scores for all subsequences as described by the following quantum steps. 

\textit{Step 3.1} Add an ancilla register and apply $H$ gate on Eq. (21) to create
\begin{equation}
\frac{1}{K}\sum_{i=1}^K|i\rangle\frac{1}{\sqrt{2}}(|0\rangle+|1\rangle)\sum_{k=1}^K|k\rangle|\bar{S}_{\mu}(X_i,X_k)\rangle|0\rangle.
\end{equation}

\textit{Step 3.2} Perform unitary operation $I\otimes|0\rangle\langle0|\otimes U+I\otimes|1\rangle\langle1|\otimes I$, where $U$ is a controlled rotation operation, it rotates $|0\rangle$ to $\xi_{i,k}|0\rangle+\sqrt{1-\xi_{i,k}^2}|1\rangle$ conditioned on $|\bar{S}_{\mu}(X_i,X_k)\rangle$, where $\xi_{i,k}=\bar{S}_{\mu}(X_i,X_k)$. Undo the fourth register to obtain
\begin{align}
&\frac{1}{K}\sum_{i=1}^K|i\rangle\frac{1}{\sqrt{2}}\big[|0\rangle\sum_{k=1}^K|k\rangle(\xi_{i,k}|0\rangle+\sqrt{1-\xi_{i,k}^2}|1\rangle)+\nonumber\\
&|1\rangle\sum_{k=1}^K|k\rangle|0\rangle\big]:=\frac{1}{\sqrt{K}}\sum_{i=1}^K|i\rangle\frac{1}{\sqrt{2}}[|0\rangle|\phi_i\rangle+|1\rangle|\rho\rangle],
\end{align}
where $|\phi_i\rangle=\frac{1}{\sqrt{K}}\sum_{i=1}^K|k\rangle(\xi_{i,k}|0\rangle+\sqrt{1-(\xi_{i,k}^2}|1\rangle)$ and $|\rho\rangle=\frac{1}{\sqrt{K}}\sum_{i=1}^K|k\rangle|0\rangle$.

\textit{Step 3.3} Applying the Inner Products Estimation (Lemma 3.2) to generate
\begin{equation}
\frac{1}{\sqrt{K}}\sum_{i=1}^K|i\rangle|\langle\phi_i|\rho\rangle\rangle=\frac{1}{\sqrt{K}}\sum_{i=1}^K|i\rangle|\sum_{k=1}^K\frac{\bar{S}_{\mu}(X_i,X_k)}{K}\rangle.
\end{equation}

\textit{Step 3.4} According to step 2, we add two additional registers on Eq. (21) to prepare the state
\begin{equation}
|0\rangle\frac{1}{K}\sum_{i,k=1}^K|i\rangle|k\rangle|\bar{S}_{\mu}(X_i,X_k)\rangle|0\rangle.
\end{equation}

\textit{Step 3.5} Apply $H$ gate on the first register to get
\begin{equation}
\frac{1}{\sqrt{2}}(|0\rangle+|1\rangle)\frac{1}{K}\sum_{i,k=1}^K|i\rangle|k\rangle|\bar{S}_{\mu}(X_i,X_k)\rangle|0\rangle.
\end{equation}

\textit{Step 3.6} Perform unitary operation $|0\rangle\langle0|\otimes U+|1\rangle\langle1|\otimes I$ on Eq. (26), where $U$ is a controlled rotation operation, it rotates $|0\rangle$ to $\xi_{i,k}|0\rangle+\sqrt{1-\xi_{i,k}^2}|1\rangle$. We can obtain
\begin{align}
&\frac{1}{\sqrt{2}K}\big[|0\rangle\sum_{i,k=1}^K|i\rangle|k\rangle(\xi_{i,k}|0\rangle+\sqrt{1-\xi_{i,k}^2}|1\rangle)\nonumber\\
&+|1\rangle\sum_{i,k=1}^K|i\rangle|k\rangle|0\rangle\big]:=\frac{1}{\sqrt{2}}[|0\rangle|\Psi\rangle+|1\rangle|\Phi\rangle],
\end{align}
where $|\Psi\rangle=\frac{1}{K}\sum_{i,k=1}^K|i\rangle|k\rangle(\xi_{i,k}|0\rangle+\sqrt{1-\xi_{i,k}^2}|1\rangle), |\Phi\rangle=\frac{1}{K}\sum_{i,k=1}^K|i\rangle|k\rangle|0\rangle$ and $\xi_{i,k}=\bar{S}_{\mu}(X_i,X_k)$.

\textit{Step 3.7} Perform the Inner Product Estimation (Lemma 3.2) to get
\begin{equation}
|\langle\Psi|\Phi\rangle\rangle=|\sum_{i=1}^K\sum_{k=1}^K\frac{\bar{S}_{\mu}(X_i,X_k)}{K^2}\rangle.
\end{equation}
According to Eq.~(24), we can obtain the quantum state
\begin{equation}
\frac{1}{\sqrt{K}}\sum_{i=1}^K|i\rangle|\sum_{k=1}^K\frac{\bar{S}_{\mu}(X_i,X_k)}{K}\rangle|\sum_{i=1}^K\sum_{k=1}^K\frac{\bar{S}_{\mu}(X_i,X_k)}{K^2}\rangle.
\end{equation}

\textit{Step 3.8} Add an ancilla register and apply the QMA gate on Eq.~(29), uncompute the second and third registers to obtain
\begin{equation}
\frac{1}{\sqrt{K}}\sum_{i=1}^K|i\rangle|\frac{\frac{1}{K}\sum_{k=1}^K\bar{S}_{\mu}(X_i,X_k)}{\frac{1}{K^2}\sum_{i,k=1}^K\bar{S}_{\mu}(X_i,X_k)}\rangle=\frac{1}{\sqrt{K}}\sum_{i=1}^K|i\rangle|h_i\rangle,
\end{equation} where $\frac{\frac{1}{K}\sum_{k=1}^K\bar{S}_{\mu}(X_i,X_k)}{\frac{1}{K^2}\sum_{i,k=1}^K\bar{S}_{\mu}(X_i,X_k)}=\frac{\frac{K}{\sqrt{q}}\sum_{k=1}^K\sqrt{\sum_{t=1}^q(\frac{\mu_i^t-\mu_k^t}{2C})^2}}{\frac{1}{\sqrt{q}}\sum_{i,k=1}^K\sqrt{\sum_{t=1}^q(\frac{\mu_i^t-\mu_k^t}{2C})^2}}$\\
$=\frac{\sum_{k=1}^KS_{\mu}(X_i,X_k)/K}{\sum_{i,k=1}^KS_{\mu}(X_i,X_k)/K^2}=h_i$.

\textit{Step 4} Grover's algorithm is applied to find all indices $i$ of the abnormal subsequences that satisfy $h_i\geq\delta$.

\subsection{Complexity and Error Analysis}

(i) We analyze the complexity of step 1. The complexity of this step is $O(\frac{\sqrt{n}\log (Kqn)}{\varepsilon_1})$, the specific analysis is as follows:

In steps 1.2-1.4, it takes $H$ and QMA gates, the oracle $O_{X}$ and $O_s$ with complexity $\log(Kqn)$. In step 1.5, the complexity of amplitude amplification is $O(\sqrt{n}\log (Kqn))$. 
In step 1.6, it contains a controlled rotation operator with complexity $O(1)$. In step 1.7, the amplitude estimation block needs $O(1/\varepsilon_1)$ applications of $Q_i^t$ to achieve error $\varepsilon_1$, and the complexity of performing unitary operator $Q_i^t$ is $O[\sqrt{n}\log (Kqn)\ln(1/\varepsilon_1)]$. In step 1.8, it takes QMA and sine gates with complexity $O($poly$\log(1/\varepsilon_1))$, which is smaller than $O(1/\varepsilon_1)$, the complexity of these gates can be omitted \cite{STJ2017}.

Now, we analyze the error of $\mu_{i}^t$, which mainly comes from the amplitude estimation in step 1.7,
\begin{align}
|\widehat{\mu}_i^t-\mu^t_i|&=C|sin^2\widehat{\theta}_i^t-sin^2{\theta}_i^t|\le|\sin(\widehat{\theta}_i^t-\theta_i^t)|\nonumber\\
&\le C\cdot|\widehat{\theta}_i^t-\theta_i^t|\le C\varepsilon_1,
\end{align}
where $\widehat{\mu}_i^t$ represents the estimate of $\mu_i^t$ and $|\widehat{\theta}_i^t-\theta_i^t|\le\varepsilon_1$ comes from step 1.8. For convenience, we use $\widehat{b}$ to denote
the estimation of $b$ in the following sections.

(ii) The time complexity of step 2 is $O(\frac{\sqrt{n}\log (Kqn)}{\varepsilon_1\varepsilon_2})$, which mainly stems from the amplitude estimation of step 2.4.

In step 2.1, the operations of step 1 are performed with complexity $O(\frac{\sqrt{n}\log (Kqn)}{\varepsilon_1})$. In steps 2.2-2.3, QMA gate and controlled rotation are performed with complexity $O($poly$\log(1/\varepsilon_1))$, which can be ignored. In step 2.4, the amplitude estimation costs $O(\frac{\sqrt{n}\log (Kqn)}{\varepsilon_1\varepsilon_2})$ time to ensure the error $\varepsilon_2$.

The error of step 2.4 is $|\widehat{\bar{S}}_{\mu}(X_i,X_k)-\bar{S}_{\mu}(X_i,X_k)|\le|\widehat{\alpha}_{i,k}-\alpha_{i,k}|\le\varepsilon_2$. We analyze the error of $\bar{S}_{\mu}(X_i,X_k)$ as follow:
\begin{align}
&|\widehat{\bar{S}}_{\mu}(X_i,X_k)-\frac{1}{\sqrt{q}}\sqrt{\sum_{t=1}^q(\frac{\mu_i^t-\mu_k^t}{2C})^2}|\nonumber\\
&\le\varepsilon_2+|\bar{S}_{\mu}(X_i,X_k)-\frac{1}{\sqrt{q}}\sqrt{\sum_{t=1}^q(\frac{\mu_i^t-\mu_k^t}{2C})^2}|,
\end{align}
where $\bar{S}_{\mu}(X_i,X_k)=\frac{1}{\sqrt{q}}\sqrt{\sum_{t=1}^q(\frac{\widehat{\mu}_i^t-\widehat{\mu}_k^t}{2C})^2}$. We assume that at least half of the values of $|\frac{\mu_i^t-\mu_k^t}{2C}|,~(t=1,2,\cdots,q)$ are greater than a constant $E$, i.e., $\bar{S}_{\mu}(X_i,X_k)\ge\frac{\sqrt{2}}{2}E,~\frac{1}{\sqrt{q}}\sqrt{\sum_{t=1}^q(\frac{\mu_i^t-\mu_k^t}{2C})^2}\ge\frac{\sqrt{2}}{2}E$. The second term of Eq. (32) is as follows:
\begin{align}
&\big|\bar{S}_{\mu}(X_i,X_k)-\frac{1}{\sqrt{q}}\sqrt{\sum_{t=1}^q(\frac{\mu_i^t-\mu_k^t}{2C})^2}|\nonumber\\
&=|\frac{\bar{S}^2_{\mu}(X_i,X_k)-\frac{1}{q}\sum_{t=1}^q(\frac{\mu_i^t-\mu_k^t}{2C})^2}{\bar{S}_{\mu}(X_i,X_k)+\frac{1}{\sqrt{q}}\sqrt{\sum_{t=1}^q(\frac{\mu_i^t-\mu_k^t}{2C})^2}}\big|\nonumber\\
&\le|\frac{\frac{1}{q}\sum_{t=1}^q[(\frac{\widehat{\mu}_i^t-\widehat{\mu}_k^t}{2C})^2-(\frac{\mu_i^t-\mu_k^t}{2C})^2]}{\sqrt{2}E}|\nonumber\\
&\le|\frac{\frac{1}{q}\sum_{t=1}^q2\cdot\frac{(\widehat{\mu}_i^t-\widehat{\mu}_k^t)-(\mu_i^t-\mu_k^t)}{2C}}{E}|\le\frac{2\varepsilon_1}{E}.
\end{align}
Therefore, we get $\bar{S}_{\mu}(X_i,X_k)$ with error $\varepsilon_2+\frac{2\varepsilon_1}{E}$.

(iii) The complexity of step 3 is $O(\frac{\sqrt{n}\log (Kqn)}{\varepsilon_1\varepsilon_2\varepsilon_3}+\frac{\sqrt{n}\log (Kqn)}{\varepsilon_1\varepsilon_2\varepsilon_4})$, which mainly coming from the Inner Product Estimation of steps 3.3 and 3.7.

In steps 3.1-3.2, $H$ gate and unitary operation $I\otimes|0\rangle\langle0|\otimes U+I\otimes|1\rangle\langle1|\otimes I$ are performed with complexity $O($poly$\log(1/\varepsilon_2))$. According to Lemma 3.3, the state $|\langle\phi_i|\rho\rangle\rangle$ can be prepared with error $\varepsilon_3$, so the complexity of step 3.3 is $O(\frac{\sqrt{n}\log (Kqn)}{\varepsilon_1\varepsilon_2\varepsilon_3})$. In step 3.4, the operations of step 2 are performed with complexity $O(\frac{\sqrt{n}\log (Kqn)}{\varepsilon'\varepsilon''})$. The complexity of steps 3.5 and 3.6 is $O($poly$\log(1/\varepsilon_2))$. In step 3.7, the Inner Product Estimation is performed with complexity $O(\frac{\sqrt{n}\log (Kqn)}{\varepsilon_1\varepsilon_2\varepsilon_4})$. The complexity of step 1, step 2 and step 3 are illustrated in TABLE I.
\begin{table}[htbp]
	\captionsetup{singlelinecheck=off}
	\caption{The time complexity of step 1, step 2 and step 3}
	\begin{tabular}{ccccccc}
		\hline\hline
		steps&complexity&steps&complexity\\ \hline
		1.1-1.4&$O(\log (Kqn))$&2.1&$O(\frac{\sqrt{n}\log(Kqn)}{\varepsilon_1})$\\
		$1.5$&$O(\sqrt{n}\log(Kqn))$ &2.2&$O($poly$\log(1/\varepsilon_1))$\\
		$1.7$& $O(\frac{\sqrt{n}\log(Kqn)}{\varepsilon_1})$&2.3&$O($poly$\log(1/\varepsilon_1))$\\
		1.8&$O($poly$\log(1/\varepsilon_1))$ &2.4&$O(\frac{\sqrt{n}\log(Kqn)}{\varepsilon_1\varepsilon_2})$\\ \hline
		3.1-3.2&$O($poly$\log(1/\varepsilon_2))$&3.6&$O($poly$\log(1/\varepsilon_2))$ \\
        3.3&$O(\frac{\sqrt{n}\log(Kqn)}{\varepsilon_1\varepsilon_2\varepsilon_3})$&3.7&$O(\frac{\sqrt{n}\log(Kqn)}{\varepsilon_1\varepsilon_2\varepsilon_4})$ \\
        3.4&$O(\frac{\sqrt{n}\log(Kqn)}{\varepsilon_1\varepsilon_2})$&3.8&$O(\log(\frac{1}{\varepsilon_3})+\log(\frac{1}{\varepsilon_4}))$ \\
        3.5&$O($poly$\log(1/\varepsilon_2))$&$~$&$~$ \\ \hline\hline
	\end{tabular}
\end{table}

Now, we analyze the errors of $\sum_{k=1}^K\bar{S}_{\mu}(X_i,X_k)/K$ and $\sum_{i,k=1}^K\bar{S}_{\mu}(X_i,X_k)/K^2$ as follow:
\begin{align}
&\bigg|\langle\phi_i\widehat{|}\rho\rangle-\frac{1}{K}\sum_{k=1}^K\frac{1}{\sqrt{q}}\sqrt{\sum_{t=1}^q(\frac{\mu_i^t-\mu_k^t}{2C})^2}\bigg|\nonumber\\
&=\bigg|\langle\phi_i\widehat{|}\rho\rangle-\langle\phi_i|\rho\rangle+\langle\phi_i|\rho\rangle-\frac{1}{K\sqrt{q}}\sum_{k=1}^K\sqrt{\sum_{t=1}^q(\frac{\mu_i^t-\mu_k^t}{2C})^2}\bigg|\nonumber\\
&\le\varepsilon_3+\bigg|\langle\phi_i|\rho\rangle-\frac{1}{K\sqrt{q}}\sum_{k=1}^K\sqrt{\sum_{t=1}^q(\frac{\mu_i^t-\mu_k^t}{2C})^2}\bigg|\nonumber\\
&\le\varepsilon_3+\varepsilon_2+\frac{2\varepsilon_1}{E},
\end{align}
where $\langle\phi_i|\rho\rangle=\sum_{k=1}^K\widehat{\bar{S}}_{\mu}(X_i,X_k)/K$ and $\bar{S}_{\mu}(X_i,X_k)=\frac{1}{\sqrt{q}}\sqrt{\sum_{t=1}^q(\frac{\widehat{\mu}_i^t-\widehat{\mu}_k^t}{2C})^2}$.
\begin{align}
&\bigg|\langle\Psi\widehat{|}\Phi\rangle-\frac{1}{K^2\sqrt{q}}\sum_{i,k=1}^K\sqrt{\sum_{t=1}^q(\frac{\mu_i^t-\mu_k^t}{2C})^2}\bigg|\nonumber\\
&=\bigg|\langle\Psi\widehat{|}\Phi\rangle-\langle\Psi|\Phi\rangle+\langle\Psi|\Phi\rangle-\frac{1}{K^2\sqrt{q}}\sum_{i,k=1}^K\sqrt{\sum_{t=1}^q(\frac{\mu_i^t-\mu_k^t}{2C})^2}\bigg|\nonumber\\
&\le\varepsilon_4+\bigg|\langle\Psi|\Phi\rangle-\frac{1}{K^2\sqrt{q}}\sum_{i,k=1}^K\sqrt{\sum_{t=1}^q(\frac{\mu_i^t-\mu_k^t}{2C})^2}\bigg|\nonumber\\
&\le\varepsilon_4+\varepsilon_2+\frac{2\varepsilon_1}{E},
\end{align}
where $\langle\Psi|\Phi\rangle=\sum_{i=1}^K\sum_{k=1}^K\widehat{\bar{S}}_{\mu}(X_i,X_k)/K^2$.
Finally, we analyze the error of $h_i$ as follows:
\begin{align}
|\widehat{h}_i-h_i|&=\big|\frac{\langle\phi_i\widehat{|}\rho\rangle}{\langle\Psi\widehat{|}\Phi\rangle}-\frac{\frac{1}{K\sqrt{q}}\sum_{k=1}^K\sqrt{\sum_{t=1}^q(\frac{\mu_i^t-\mu_k^t}{2C})^2}}{\frac{1}{K^2\sqrt{q}}\sum_{i,k=1}^K\sqrt{\sum_{t=1}^q(\frac{\mu_i^t-\mu_k^t}{2C})^2}}\big|\nonumber\\
&\le\big|\frac{\langle\phi_i\widehat{|}\rho\rangle-\frac{1}{K\sqrt{q}}\sum_{k=1}^K\sqrt{\sum_{t=1}^q(\frac{\mu_i^t-\mu_k^t}{2C})^2}}{E}\big|\nonumber\\
&\le\frac{(\varepsilon_3+\varepsilon_2)}{E}+\frac{2\varepsilon_1}{E^2}.
\end{align}

If $\varepsilon_1=\frac{E^2\varepsilon}{6},~\varepsilon_2=\frac{E\varepsilon}{3}, \varepsilon_3=\frac{E\varepsilon}{3}$, and $\varepsilon_4=\varepsilon$, we can get $|\widehat{h}_i-h_i|\le\varepsilon$. That is, we can obtain an $\varepsilon$-approximate of the state $\frac{1}{\sqrt{K}}\sum_{i=1}^K|i\rangle|h_i\rangle$ with complexity $O(\frac{E^3\sqrt{n}\log(Kqn)}{\varepsilon^3})$ in step 3.

(iv) The complexity of executing Grover's algorithm is $O(\sqrt{KT}\cdot\frac{E^3\sqrt{n}\log(Kqn)}{\varepsilon^3})$ to obtain all abnormal subsequences, where $T$ is the number of abnormal subsequences. It imply that the complexity of obtaining all anomalous subsequences is $O(\sqrt{KT}\cdot\frac{E^3\sqrt{n}\log(Kqn)}{\varepsilon^3})$.

We know that $E=O(1)$, and in general the number $T$ of abnormal subsequences is much smaller than the number $K$ of subsequences, the overall runtime will be $O(\frac{\sqrt{Kn}\log(Kqn)}{\varepsilon^3})$. Our quantum algorithm achieves polynomial speedup compared to its classical counterpart.

\section{Conclusion}
\label{sec4}
In practical application scenarios, due to the huge amount of data and the difficulty of collecting abnormal labels or normal label samples, the data is often unlabeled. This requires unsupervised anomaly detection to identify anomalies. The cost of performing classical ADPAAD is too much when dealing with large-scale sequences, we proposed a quantum algorithm for ADPAAD. It achieves a polynomial speedup compared with its classical counterpart. Furthermore, our proposed quantum algorithm for PAAD representation of each subsequence can be reused as a subroutine for other quantum algorithms.

Our approach opens an avenue for the quantum algorithm of  unsupervised anomaly detection. Designing more quantum algorithms for unsupervised anomaly detection is a goal worth considering in the future. We hope our algorithm can inspire more efficient quantum anomaly detection algorithms.

\section*{Acknowledgements}
We thank Di Zhang, Yongmei Li, Jing Li, and Zhenqiang Li for useful discussions on the subject. This work is supported by NSFC (Grants Nos.61976024, 61972048, 62072051).
\appendix
\section{Detailed analysis of the general case of step 1}
\label{sec6}
In the general case, according to Ref. \cite{SLH2022}, we perform amplitude amplification in step 1.5 to obtain the quantum state
\begin{equation}
\frac{1}{\sqrt{K}}\sum_{i=1}^K|i\rangle_1\frac{1}{\sqrt{q}}\sum_{t=1}^{q}|t\rangle_2(\sqrt{p}|\Phi_i^t\rangle+\sqrt{1-p}|(\Phi_i^t)^{\perp}\rangle)_{567},
\end{equation}
where $|\Phi_i^t\rangle=\frac{1}{\sqrt{n_{i}^{t}}}\sum_{x_{i}(j)\in I_{i}^{t}}|j\rangle|x_i(j)\rangle|\rho_{i}^{t}(j)\rangle$, $n_{i}^t$ denotes the number of data points belonging to the subsection $I_{i}^t$ in subsequence $X_{i}$, $p$ represents the probability of successfully measuring $|\Phi_i^t\rangle$, and $|(\Phi_i^t)^{\perp}\rangle$ is the quantum state that is orthogonal to $|\Phi_i^t\rangle$.

Then we perform a controlled operation in step (1.6), which satisfies that when $\rho_i^t(j)\le0$ in the seventh register, a controlled rotation is performed, and when $\rho_i^t(j)>0$, an $XOR$ gate is performed on the  ancilla register. We can get
\begin{align}
&\frac{1}{\sqrt{qK}}\sum_{i=1}^K|i\rangle_1\sum_{t=1}^{q}|t\rangle_2\big[\sqrt{p}|\Phi_i^t\rangle(\sqrt{\frac{x_{i}(j)}{C}}|0\rangle+\sqrt{1-\frac{x_{i}(j)}{C}}|1\rangle)\nonumber\\
&+\sqrt{1-p}|(|\Phi_i^t)^{\perp}\rangle|1\rangle]_{5678}:=\frac{1}{\sqrt{qK}}\sum_{i=1}^K|i\rangle_1\sum_{t=1}^{q}|t\rangle_2|\psi_i^t\rangle_{5678}.
\end{align}
The $|\psi_i^t\rangle$ can be rewritten as $\sin\theta_i^t|(\psi_i^t)^0\rangle+\cos\theta_i^t|(\psi_i^t)^1\rangle$, where $|(\psi_i^t)^0\rangle$ denotes the normalized quantum state of $|\Phi_i^t\rangle\sqrt{\frac{x_{i}(j)}{C}}|0\rangle$ and $|(\psi_i^t)^1\rangle$ is the quantum state that is orthogonal to $|(\psi_i^t)^0\rangle$. We can use amplitude estimation to estimate $\sin^2(\theta_{i}^{t})=\frac{p}{n_{i}^t}\sum_{x_{i}(j)\in I_{i}^t}\frac{x_{i}(j)}{C}$.

By performing the operations of steps (1.7)-(1.8), we can obtain
\begin{equation}
\frac{1}{\sqrt{qK}}\sum_{i=1}^K|i\rangle\sum_{t=1}^{q}|t\rangle|\bar{\mu}_i^t\rangle=\frac{1}{\sqrt{qK}}\sum_{i=1}^K|i\rangle\sum_{t=1}^{q}|t\rangle|C\cdot\sin^2(\theta_{i}^{t})\rangle,
\end{equation}
where $\bar{\mu}_i^t=p\cdot\mu_i^t=C\cdot\sin^2(\theta_{i}^{t})$.

Our analysis of the general case of step 1 does not affect the subsequent process and final result of the quantum algorithm, that is, we can perform the operations of steps 2 and 3 to get
\begin{equation}
\frac{1}{\sqrt{K}}\sum_{i=1}^K|i\rangle|\frac{K\sum_{k=1}^K\bar{S}_{\mu}(X_i,X_k)}{\sum_{i=1}^K\sum_{k=1}^K\bar{S}_{\mu}(X_i,X_k)}\rangle=\frac{1}{\sqrt{K}}\sum_{i=1}^K|i\rangle|h_i\rangle,
\end{equation}
\begin{equation}
\bar{S}_{\mu}(X_i,X_k)=\frac{1}{\sqrt{q}}\sqrt{\sum_{t=1}^q(\frac{\bar{\mu}_i^t-\bar{\mu}_k^t}{2C})^2}=\frac{p}{\sqrt{q}}\sqrt{\sum_{t=1}^q(\frac{\mu_i^t-\mu_k^t}{2C})^2}.\nonumber
\end{equation}

We can still get
\begin{align}
&\frac{\frac{1}{K}\sum_{k=1}^K\bar{S}_{\mu}(X_i,X_k)}{\frac{1}{K^2}\sum_{i=1}^K\sum_{k=1}^K\bar{S}_{\mu}(X_i,X_k)}=\frac{\frac{Kp}{\sqrt{q}}\sum_{k=1}^K\sqrt{\sum_{t=1}^q(\frac{\mu_i^t-\mu_k^t}{2C})^2}}{\frac{p}{\sqrt{q}}\sum_{i,k=1}^K\sqrt{\sum_{t=1}^q(\frac{\mu_i^t-\mu_k^t}{2C})^2}}\nonumber\\
&=\frac{\sum_{k=1}^KS_{\mu}(X_i,X_k)/K}{\sum_{i,k=1}^KS_{\mu}(X_i,X_k)/K^2}=h_i.
\end{align}


\begin{thebibliography}{0}%
\makeatletter
\providecommand \@ifxundefined [1]{%
 \@ifx{#1\undefined}
}%
\providecommand \@ifnum [1]{%
 \ifnum #1\expandafter \@firstoftwo
 \else \expandafter \@secondoftwo
 \fi
}%
\providecommand \@ifx [1]{%
 \ifx #1\expandafter \@firstoftwo
 \else \expandafter \@secondoftwo
 \fi
}%
\providecommand \natexlab [1]{#1}%
\providecommand \enquote  [1]{``#1''}%
\providecommand \bibnamefont  [1]{#1}%
\providecommand \bibfnamefont [1]{#1}%
\providecommand \citenamefont [1]{#1}%
\providecommand \href@noop [0]{\@secondoftwo}%
\providecommand \href [0]{\begingroup \@sanitize@url \@href}%
\providecommand \@href[1]{\@@startlink{#1}\@@href}%
\providecommand \@@href[1]{\endgroup#1\@@endlink}%
\providecommand \@sanitize@url [0]{\catcode `\\12\catcode `\$12\catcode
  `\&12\catcode `\#12\catcode `\^12\catcode `\_12\catcode `\%12\relax}%
\providecommand \@@startlink[1]{}%
\providecommand \@@endlink[0]{}%
\providecommand \url  [0]{\begingroup\@sanitize@url \@url }%
\providecommand \@url [1]{\endgroup\@href {#1}{\urlprefix }}%
\providecommand \urlprefix  [0]{URL }%
\providecommand \Eprint [0]{\href }%
\providecommand \doibase [0]{http://dx.doi.org/}%
\providecommand \selectlanguage [0]{\@gobble}%
\providecommand \bibinfo  [0]{\@secondoftwo}%
\providecommand \bibfield  [0]{\@secondoftwo}%
\providecommand \translation [1]{[#1]}%
\providecommand \BibitemOpen [0]{}%
\providecommand \bibitemStop [0]{}%
\providecommand \bibitemNoStop [0]{.\EOS\space}%
\providecommand \EOS [0]{\spacefactor3000\relax}%
\providecommand \BibitemShut  [1]{\csname bibitem#1\endcsname}%
\let\auto@bib@innerbib\@empty
\end{thebibliography}%


\begin{thebibliography}{}
	
\bibitem{VAV2009} V. Chan-dola, A. Baner-jee, V. Ku-mar, A. Val-aba, Anomaly detection: A survey, ACM computing surveys (CSUR) 41.3, 1-58 (2009).

\bibitem{RLL2017} H R. Ren, X J. Liao, Z W. Li, A. AI-Ahmari, Anomaly detection using piecewise aggregate approximation in the amplitude domain, Applied Intelligence (2017).

\bibitem{TTP2011} D K. Tewatia, R P. Tolakanahalli, B R. Paliwal, Time series analyses of breathing patterns of lung cancer patients using nonlinear dynamical system theory, Phys Med Biol 56.7, 2161 (2011).

\bibitem{VDM2009} J. Viinikka, H. Debar, M\'e. Ludovic, A. Lehikoinen, M. Tarvainen, Processing intrusion detection alert aggregates with time series modeling, Information Fusion 10.4, 312-324 (2009).

\bibitem{CFL2011} P C. Chang, C Y. Fan, J L. Lin, Trend discovery in financial time series data using a case based fuzzy decision tree, Expert Systems with Applications 38.5, 6070每6080 (2011).

\bibitem{ADK2010} M. Avazbeigi, S. Doulabi, B. Karimi, Choosing the appropriate order in fuzzy time series: A new N-factor fuzzy time series for prediction of the auto industry production, Expert Systems with Applications 37.8, 5630每5639 (2010).

\bibitem{LBF2013} M. Lippi, M. Bertini, P. Frasconi, Short-Term Traffic Flow Forecasting: An Experimental Comparison of Time-Series Analysis and Supervised Learning, IEEE Transactions on Intelligent Transportation Systems 14.2, 871每882 (2013).


\bibitem{PW1994} P W. Shor, Algorithms for quantum computation: Discrete logarithms and factoring, In Proceedings of 35th Annual Symposium on the Foundations of Computer Science, IEEE Computer Society Press, Los Alamitos, CA 124-134 (1994).

\bibitem{LK1996} L K. Grover, A fast quantum mechanical algorithm for database search, In Proceedings of the twenty-eighth annual ACM symposium on Theory of computing 212-219 (1996).

\bibitem{AAS2009} A W. Harrow, A. Hassidim, S. Lloyd, Quantum algorithm for linear systems of equations, Phys. Rev. Lett. 103.15, 150502 (2009).

\bibitem{LCS2018} L C. Wan, C H. Yu, S J. Pan, F. Gao, Q Y. Wen, S J. Qin, Asymptotic quantum algorithm for the Toeplitz systems, Phys. Rev. A 97.6, 062322 (2018).

\bibitem{HYW2021} H L. Liu, Y S. Wu, L C. Wan, S J. Pan, F. Gao, S J. Qin, Q Y. Wen, Variational quantum algorithm for the Poisson equation, Phys. Rev. A 104.2, 022418 (2021).

\bibitem{ZBH2022} Z Q. Li, B B. Cai, H W. Sun, H L. Liu, L C. Wan, S J. Qin, Q Y. Wen, F. Gao, Novel quantum circuit implementation of Advanced Encryption Standard with low costs, Sci. China Phys. Mech. Astron. 65, 290311 (2022).

\bibitem{XZX2019} X Y. Dong, Z. Li, X Y Wang, Quantum cryptanalysis on some generalized Feistel schemes, Science China Information Sciences 62.2, 1-12 (2019).

\bibitem{CXT2020} C Y. Wei, X Q. Cai, T Y. Wang, S J. Qin, F. Gao, Q Y. Wen, Error Tolerance Bound in QKD-Based Quantum Private Query, IEEE Journal on Selected Areas in Communications 38, 517-527 (2020).

\bibitem{FSW2019} F. Gao, S J. Qin, W. Huang, Q Y. Wen, Quantum private query: a new kind of practical quantum cryptographic protocols, Sci. China-Phys. Mech. Astron. 62, 070301 (2019).

\bibitem{GLM2008} V. Giovannetti, S. Lloyd, L. Maccone, Quantum Private Queries, Phys. Rev. Lett. 100.23, 230502 (2008).

\bibitem{SMP2013} S. Lloyd, M. Mohseni, P. Rebentrost, Quantum algorithms for supervised and unsupervised machine learning, arXiv:1307.0411 (2013).

\bibitem{NDS2012} N. Wiebe, D. Braun, S. Lloyd, Quantum algorithm for data fitting, Phys. Rev. Lett. 109.5, 050505 (2012).

\bibitem{SBS2017} S C. Morampudi, B. Hsu, S L. Sondhi, R. Moessner, Clustering in Hilbert space of a quantum optimization problem, Phys. Rev. A 96.4, 042303 (2017).

\bibitem{PTC2018} P. Rebentrost, T R. Bromley, C. Weedbrook, S. Lloyd, Quantum Hopfield neural network, Phys. Rev. A 98.4, 042308 (2018).

\bibitem{GM2017} G M. Wang, Quantum algorithm for linear regression, Phys. Rev. A 96.1, 012335 (2017).

\bibitem{CFQ2019} C H. Yu, F. Gao, Q Y. Wen, An improved quantum algorithm for ridge regression, IEEE Transactions on Knowledge and Data Engineering (2019).

\bibitem{CFC2019} C H. Yu, F. Gao, C. Liu, D. Huynh, M. Reynolds, J. Wang, Quantum algorithm for visual tracking, Phys. Rev. A 99.2, 022301 (2019).

\bibitem{CFQ2016} C H. Yu, F. Gao, Q L. Wang, Q Y. Wen, Quantum algorithm for association rules mining, Phys. Rev. A 94.4, 042311 (2016).

\bibitem{IL2016} I. Cong, L. Duan, Quantum discriminant analysis for dimensionality reduction and classification, New J. Phys. 18, 073011 (2016).

\bibitem{SMP2014} S. Lloyd, M. Mohseni, P. Rebentrost, Quantum principal component analysis, Nature Physics 10, 631 (2014).

\bibitem{SLH2020} S J. Pan, L C.Wan, H L. Liu, F. Gao, S J. Qin, Q Y.Wen, Improved quantum algorithm for A-optimal projection, Phys. Rev. A 102.5, 052402 (2020).

\bibitem{CFS2019} C H. Yu, F. Gao, S. Lin, J. Wang, Quantum data compression by principal component analysis, Quantum Information Processing 18.8, 1-20 (2019).

\bibitem{PMS2014} P. Rebentrost, M. Mohseni, S. Lloyd, Quantum support vector machine for big data classification, Phys. Rev. Lett. 113.13, 130503 (2014).

\bibitem{ZLH2020} Z K. Ye, L Z. Li, H Z. Situ, Y Y. Wang, Quantum speedup for twin support vector machines, Sci. China Inf. Sci. 63.8, 189501 (2020).

\bibitem{NP2018} N. Liu, P. Rebentrost, Quantum machine learning for quantum anomaly detection, Phys. Rev. A 97.4, 042315 (2018).

\bibitem{JSM2019} J M. Liang, S Q. Shen, M. Li, L. Li, Quantum anomaly detection with density estimation and multivariate Gaussian distribution, Phys. Rev. A 99.5, 052310 (2019).

\bibitem{MHY2021} M C. Guo, H L. Liu, Y M. Li, W M. Li, F. Gao, S J. Qin, Q Y. Wen, Quantum algorithms for anomaly detection using amplitude estimation, Physica A: Statistical Mechanics and its Applications 604, 127936 (2022).

\bibitem{LJ2017} L. Ruiz-Perez, J C. Garcia-Escartin, Quantum arithmetic with the quantum fourier transform, Quantum Information Processing 16.6, 152 (2017).

\bibitem{STJ2017} S S. Zhou, T. Loke, J A. Izaac, J B. Wang, Quantum Fourier transform in computational basis, Quantum Information Processing 16.3, 82 (2017).

\bibitem{GPM2002} G. Brassard, P. Hoyer, M. Mosca, Quantum amplitude amplification and estimation, Contemporary Mathematics 305 (2002).

\bibitem{SLH2022} S J. Pan, L C. Wan, H L. L, Y S. W, S J. Qin, Q Y. Wen, F. Gao, Quantum algorithm for neighborhood preserving embedding, Chinese Physics B 31.6, 060304 (2022).

\bibitem{VSL2008} V. Giovannetti, S. Lloyd, L. Maccone, Quantum Random Access Memory, Phys. Rev. Lett. 100.16, 160501 (2008).


\bibitem{KJA2019} K. Iordanis, J. Landman, A. Luongo, A. Prakash, q-means: A quantum algorithm for unsupervised machine learning, Advances in Neural Information Processing Systems 32 (2019).

\bibitem{BJY2018} B J. Duan, J B. Yuan, Y. Liu, D. Li, Efficient quantum circuit for singular-value thresholding, Phys. Rev. A 98.1, 012308 (2018).

\bibitem{KMK2019} K. Mitarai, M. Kitagawa, K. Fujii, Quantum analog-digital conversion, Phys. Rev. A 99.1, 012301 (2019).
\end{thebibliography}
\end{document}